\documentclass[journal=jctcce,manuscript=article]{achemso}

\usepackage[version=3]{mhchem}
\usepackage{chemformula}
\usepackage[english]{babel}
\usepackage[T1]{fontenc}
\usepackage[utf8]{inputenc}

\usepackage[para,online,flushleft]{threeparttable}
\usepackage{booktabs}
\usepackage{graphicx}
\usepackage[mathscr]{eucal}
\usepackage{mathtools}
\usepackage{bm}
\usepackage{mathptmx}
\usepackage{amsfonts}
\usepackage{amsmath}
\usepackage{amssymb}
\usepackage{enumitem}
\usepackage{float}
\usepackage{multirow}
\usepackage{comment}
\usepackage[normalem]{ulem}
\usepackage{color,soul}
\usepackage{array}
\usepackage{arydshln}

\author{Markel Ylla}
\affiliation{Donostia International Physics Center (DIPC), 20018 Donostia, Euskadi, Spain}
\affiliation{Kimika Fakultatea, Euskal Herriko Unibertsitatea UPV/EHU, P.K. 1072, 20080 Donostia, Euskadi, Spain}

\author{Xiang Xu}
\affiliation{Donostia International Physics Center (DIPC), 20018 Donostia, Euskadi, Spain}

\author{Eloy Ramos-Cordoba}
\affiliation{Institute for Advanced Chemistry of Catalonia (IQAC), CSIC, Jordi Girona 18-26, 08034 Barcelona, Spain}

\author{Eduard Matito}
\email{ematito@dipc.org}
\affiliation{Donostia International Physics Center (DIPC), 20018 Donostia, Euskadi, Spain}
\affiliation{Kimika Fakultatea, Euskal Herriko Unibertsitatea UPV/EHU, P.K. 1072, 20080 Donostia, Euskadi, Spain}
\affiliation{Ikerbasque Foundation for Science, Plaza Euskadi 5, 48009 Bilbao, Euskadi, Spain}

\title{Dynamic Correlation through Natural-Orbital Occupation Entropy}


\begin{document}


\begin{abstract}
We introduce an information-theoretical indicator of dynamic electron correlation based on the Shannon entropy of natural-orbital occupation numbers. The entropy is constructed from the normalized orbital contributions to the total correlation measure, \(I_{\mathrm{T}}\), and quantifies the dispersity of these contributions among the natural orbitals. Its exponential, \(N_{\mathrm{eff}}\), admits a direct interpretation as the effective number of orbitals contributing to the correlation measure. The proposed quantities show a strong correlation with the dynamic-correlation index \(I_{\mathrm{D}}\) for a broad set of molecular systems covering both single-reference and multireference regimes.
Beyond characterization, these measures may guide the construction of compact basis sets and accelerated wavefunction methods that exploit the locality of dynamic correlation.
\end{abstract}

\newpage
\section{Introduction}

Since the concept of electron correlation first appeared in the literature,\cite{slater:34pr} its various manifestations have been extensively investigated.\cite{kutzelnigg:68pr,cremer:01mp} Among them, the distinction between nondynamic and dynamic correlation has proven particularly useful.\cite{sinanoglu:64acp} Its importance stems from the fact that current electronic-structure methods often describe one type of correlation more effectively than the other. Because many computationally economical methods provide a satisfactory treatment of dynamic correlation for small and medium-sized molecules, a central concern in computational chemistry has been to determine whether a given molecule exhibits significant nondynamic correlation and therefore requires the use of expensive multireference methods.\cite{pulay:88jcp,lee:89tca,cioslowski:92tca,nielsen:99cpl,lee:03cpl,juhasz:06jcp,tishchenko:08jctc,alcoba:10jcp,boguslawski:12jpcl,fogueri:13tca,ramos-cordoba:16pccp,benavides:17pra,duan:20jctc,ding:21:jctc,martin:22aip,izsak:23jctc,xu:24jctc,xu:25jcp}

In contrast, considerably less attention has been paid to the role of dynamic correlation in molecular systems. Unlike nondynamic correlation, dynamic correlation is a universal feature of molecules and is present, to a greater or lesser extent, in all but a few very specific systems, which are often better regarded as physical models than as real molecules.\cite{sinanoglu:64acp,cohen:01mpb} Nevertheless, the \textit{amount} of dynamic correlation in a system is important, particularly when selecting a method from the hierarchies of wavefunction methods available in the literature, such as the MP series, CI, and CC methods, or when choosing a density functional approximation from the rungs of Jacob's ladder.\cite{tew:07jcc} In addition, measures of dynamic correlation can be employed in the construction of compact basis sets tailored to recover dynamic electron correlation,\cite{dunning:89jcp} or as proxies to exploit the \textit{locality} of dynamic correlation in the development of \textit{accelerated} wavefunction methods\cite{naim:23jctc} that exploit orbital localization\cite{pulay:83cpl,werner:11jcp,rolik:13jcp,riplinger:13jcp} or \textit{enhanced} wavefunction methods.~\cite{paulau:25jctc}

It is therefore surprising that so few dynamic-correlation measures have been proposed in the literature.\cite{cioslowski:91pra,mok:96jpc,valderrama:97jcp,ziesche:00the,valderrama:05jmc,ramos-cordoba:16pccp,ramos-cordoba:17jctc,via-nadal:19jpcl} Some possess considerable conceptual value, such as the approaches developed by Cioslowski\cite{cioslowski:91pra} and Lude\~na,\cite{valderrama:97jcp} but lack practical applicability. Others rely on comparisons with reference wavefunctions that are assumed to contain little dynamic correlation while recovering most of the nondynamic correlation; CASSCF wavefunctions are the typical choice.\cite{valderrama:05jmc,ziesche:00the,mok:96jpc} Other approaches, although conceptually appealing, require the computation of the second-order reduced density matrix.\cite{valderrama:05jmc,via-nadal:19jpcl,via-nadal:22jcp} Thus, to the best of our knowledge, there only two articles that suggest computationally economical measures of dynamic correlation, both of them based on natural-orbital occupation numbers.\cite{ziesche:00the,ramos-cordoba:16pccp}

In the first article, Ziesche introduce $I_Z$,\cite{ziesche:00the}
\begin{eqnarray}
    n_H &=& \min_{i\leq N}{\left(n_i^{\sigma}\right)}, \\
    n_L &=& \max_{i> N}\left(n_i^{\sigma}\right), \\
    I_Z &=& n_L/n_H,
\end{eqnarray}
where \(n_i^{\sigma}\) denotes the occupation number of the \(i\)th \(\sigma\)-spin natural orbital of the 1-RDM and \(n_H\) and \(n_L\) are the smallest occupied and largest virtual spin-natural-orbital occupations, respectively. For weakly correlated systems, \(I_Z\rightarrow 0\); thus, a larger deviation from zero indicates stronger correlation. However, \(I_Z\) does not distinguish well between dynamic and nondynamic correlation. Nondynamic correlation also reduces the occupation gap, so a large \(I_Z\) value may arise even when the dynamic correlation is modest. Consequently, \(I_Z\) cannot be used as a standalone measure of dynamic correlation in systems with substantial nondynamic correlation. As such, \(I_Z\) is a more useful descriptor of nondynamic correlation and is closely related to other correlation measures available in the literature.\cite{pulay:88jcp,nielsen:99cpl,tishchenko:08jctc,bartlett:20jcp,xu:24jctc,bao:25pnas} In the Supporting Information, we provide a numerical demonstration that \(I_Z\), as well as the other dynamic-correlation indices proposed by Ziesche in Ref.~\citenum{ziesche:00the}, fails to identify dynamic correlation reliably in the presence of nondynamic correlation.\newline

In 2016, we introduced a measure of dynamic electron correlation,
\begin{equation}
    I_{\mathrm{D}} =
    \frac{1}{4}\sum_{i,\sigma}
    \left[n_i^{\sigma}(1-n_i^{\sigma})\right]^{1/2}
    -\frac{1}{2}\sum_{i,\sigma}
    n_i^{\sigma}(1-n_i^{\sigma}),
\end{equation}
which can be rationalized through Lieb's
partitioning\cite{lieb:81prl,levy:87bc,buijse:91thesis}
of the pair density,\cite{via-nadal:19jpcl,via-nadal:22jcp}
as applied to a simple two-electron, two-orbital model.
This model also gives rise to measures of nondynamic and
total electron correlation,\cite{ramos-cordoba:16pccp,ramos-cordoba:17jctc}
\begin{align}
    I_{\mathrm{ND}} &=
    \frac{1}{2}\sum_{i,\sigma}
    n_i^{\sigma}(1-n_i^{\sigma}),\\
    I_{\mathrm{T}} &=
    I_{\mathrm{D}}+I_{\mathrm{ND}}
    =\frac{1}{4}\sum_{i,\sigma}
    \left[n_i^{\sigma}(1-n_i^{\sigma})\right]^{1/2}.
\end{align}

These quantities are correlation measures constructed as sums of
orbital contributions, all expressed in terms of the same
fundamental ingredient,
$x_{i}^{\sigma}=n_i^{\sigma}(1-n_i^{\sigma})$.
Their derivation from a two-electron model imposes a particular
dependence on the occupation numbers, which need not be optimal
for more general electronic systems. Nevertheless, this
dependence is reasonably well suited to $I_{\mathrm{T}}$ and
$I_{\mathrm{ND}}$, whose orbital contributions are maximal at
$n_i^{\sigma}=1/2$, consistent with the frequent occurrence
of fractionally occupied orbitals in nearly degenerate pairs.

The situation is less satisfactory for $I_{\mathrm{D}}$.
Its orbital contribution vanishes at $n_i^{\sigma}=1/2$,
a feature inherited from the two-electron model rather than
dictated by a general physical requirement for dynamic
correlation. Despite this limitation, $I_{\mathrm{D}}$
has proved to be a useful indicator of dynamic correlation, giving maximal orbital contributions of $1/32$ at
$n_i^{\sigma}=(2\pm\sqrt{3})/4
\approx 0.067,\,0.933$.\newline

The aim of the present work is to develop a more robust measure of dynamic electron correlation that retains reasonable agreement with \(I_{\mathrm{D}}\) while avoiding the arbitrary features inherited from its underlying two-electron model. Beyond its use as a tool to help assess the extent of dynamic correlation that a given wavefunction represents, this measure can guide the construction of compact basis sets tailored to recover this correlation, and assist in identifying local correlation domains for efficient electronic-structure methods.

\section{An Entropy approach to Dynamic Correlation}

Let us consider the set
$\{x_i^\sigma=n_i^\sigma(1-n_i^\sigma)\}$,
whose elements quantify the deviation of the natural orbital
occupation numbers (NOONs) from their idempotent values
of 0 and 1. In terms of $x_i^\sigma$, the correlation
indices can be expressed as
\begin{eqnarray}
    I_{\mathrm{ND}}^{\sigma} &=&
    \frac12\sum_{i}x_{i}^{\sigma}, \\
     I_{\mathrm{D}}^{\sigma} &=& \frac12
    \sum_{i}\left(x_{i}^{\sigma}\right)^{1/2}\left(\frac12-\left(x_{i}^{\sigma}\right)^{1/2}\right),\\
    I_{\mathrm{T}}^{\sigma} &=&
    \frac{1}{4}\sum_{i}\left(x_{i}^{\sigma}\right)^{1/2}.
\end{eqnarray}
The corresponding total indices are obtained by summing the spin-resolved contributions,
$I=I^\alpha+I^\beta$.\footnote{For open-shell systems, it might be more informative to consider each spin contribution rather than the sum.} For consistency, we adopt the same convention
for the indices developed in this work. Thus, the original definitions
are formulated per spin sector, whereas the corresponding total
quantities are obtained by summing the two spin-sector contributions.\newline

The individual contributions to $I_{\mathrm{T}}^\sigma$
can be used to define a normalized distribution,
\begin{eqnarray}
    r_{i}^{\sigma}=\frac{\left(x_{i}^{\sigma}\right)^{1/2}}{\sum_i \left(x_{i}^{\sigma}\right)^{1/2}},
    \hspace{2cm} \sum_i r_{i}^{\sigma} = 1,
\end{eqnarray}
which represents the relative weight of each orbital in
the total correlation measure. The Shannon entropy of
this distribution is then
\begin{equation}
    H_D^\sigma =
    -\sum_i r_i^\sigma\log r_i^\sigma,
\end{equation}
where \(\log\) denotes the natural logarithm, \(\log_e\).

For an idempotent one-particle density matrix,
all occupation numbers are either zero or one and therefore
$x_i^\sigma=0$ for every orbital. In this case, the
normalized distribution is undefined, and we adopt
$H_D^\sigma=0$ by convention.\newline

The entropy $H_D^\sigma$ quantifies the dispersity of the
orbital contributions to the total correlation measure.
Dynamic correlation is typically distributed over a large
number of natural orbitals, each making a relatively small
contribution, whereas nondynamic correlation is generally
associated with a smaller number of orbitals exhibiting
substantial deviations from integer occupations.
Consequently, dynamic correlation is expected to produce
a more broadly distributed set of orbital contributions,
and hence a larger entropy, than nondynamic correlation.
Thus, \(H_D^\sigma\) is directly related to the \textit{effective} number of natural orbitals contributing to the correlation measure \(I_T\).\newline

The minimal \textit{effective orbitals} than can couple to produce electron correlation are two, and therefore, the minimum of $H_D^\sigma$ for a correlated wavefunction is $\log(2)$.
For a distribution containing $M$ nonzero orbital contributions, $\log(2)\leq H_D^\sigma\leq\log (M)$. The lower bound is reached when the distribution includes only one \textit{donor} orbital and one \textit{acceptor} orbital, which exchange electrons numbers with respect to the canonical distribution of $N^{\sigma}$ fully occupied orbitals and $M-N^{\sigma}$ unoccupied ones, whereas the upper bound corresponds to a uniform distribution, $r_i^\sigma=1/M$, which corresponds to the strongest-correlation system that it does not occur except in idealized physical systems, such as the strongest correlated limit of the harmonium atom.\cite{cioslowski:11jctc} Thus, the entropy is sensitive not only to the number of contributing orbitals but also to the uniformity of their relative contributions. A particularly useful quantity is the exponential of the entropy,
\begin{equation}
N_{\mathrm{eff}}^\sigma=\exp(H_D^\sigma).
\end{equation}
This quantity can be interpreted as the effective number of orbitals contributing to the correlation measure. It ranges from 2 to $M$ for any nonzero distribution and equals the actual number of contributing orbitals when all their weights are identical. The exponential transformation is monotonic and therefore preserves the ordering of systems according to their entropy. Nevertheless, it provides a more intuitive representation of orbital dispersity and facilitates comparisons with additive correlation measures such
as $I_{\mathrm{D}}$ (\textit{vide infra}).\newline

An important feature of the entropy is its sensitivity to small orbital contributions. Since
\begin{equation}
\lim_{r\rightarrow0^+}(-r\log r)=0,
\qquad
\lim_{r\rightarrow0^+}
\frac{d(-r\log r)}{dr}=+\infty,
\end{equation}
weakly contributing orbitals have individually vanishing entropy contributions, although the entropy is sensitive to their presence. This behavior is particularly relevant to dynamic correlation, which typically involves a large number of orbitals with small deviations from integer occupations.\newline

$H_D^\sigma$ is not driven by the value of $I_T$ but rather by how natural orbital occupancies contribute to this quantity. Two sets of occupations that yield proportional values of $\sqrt{x_i^\sigma}$ have identical normalized distributions and therefore identical entropies.
Consequently, $H_D^\sigma$ should be interpreted as a measure of the dispersity of correlation contributions, its usefulness as a dynamic correlation indicator relying on the physical association between dynamic correlation and the participation of many weakly contributing
orbitals.\newline

\begin{figure}[H]
  \centering
  \includegraphics[width=\linewidth]{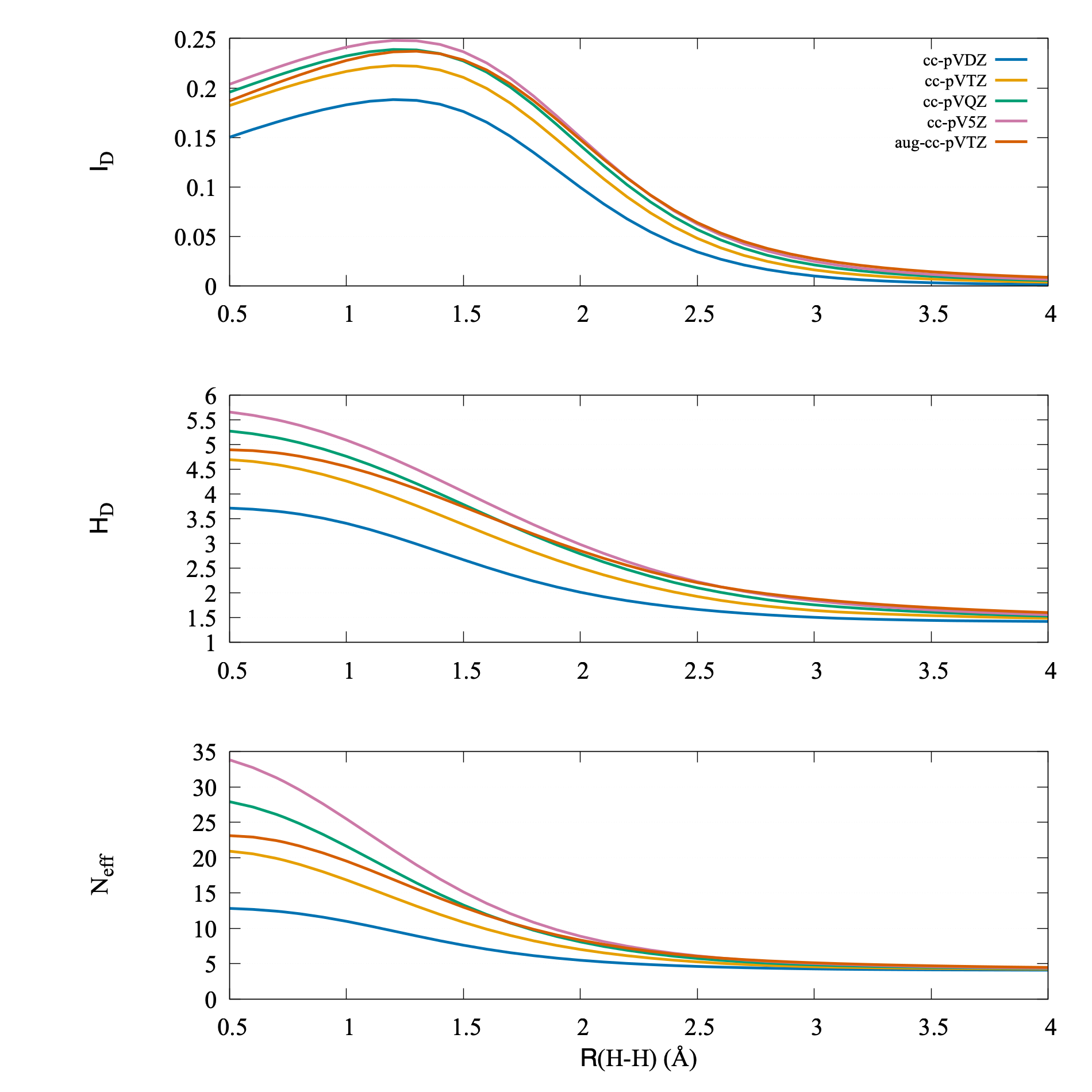}
  \caption{
    Evolution of \(I_D\) (top), \(H_D\) (middle), and \(N_{\mathrm{eff}}\) (bottom) along the dissociation of
    \(\mathrm{H}_2\). Results are shown for the cc-pVDZ, cc-pVTZ, cc-pVQZ, cc-pV5Z, and aug-cc-pVTZ basis sets in conjunction with the full-configuration interaction method.
  }
  \label{fig:h2-dissociation-id-hd}
\end{figure}

This feature should be kept in mind when employing $H_D^\sigma$ as a dynamic correlation indicator. Unless the basis set is sufficiently large, the actual value of $H_D$ will not reflect the dynamic correlation of this molecule but rather how this correlation is accounted for by the approximate wavefunction in use. Let us consider a simple illustrative example: the dissociation of $H_2$ with a minimal basis set. For each spin case, $\sigma$, since $n_1^\sigma+n_2^\sigma=1$, we have $r_1=r_2=1/2$ and thus
\begin{equation}
H_D^\sigma=
-r_1^\sigma\log r_1^\sigma
-r_2^\sigma\log r_2^\sigma=\log2,
\end{equation}
regardless the value of $n_1^\sigma$. In other words, the wavefunction does not have the basis-set flexibility to give a different distribution than $\{n_1^\sigma,1-n_1^\sigma\}$, the electron occupations distributing to the maximum of its possibilities ($N_{\mathrm{eff}}^\sigma=2$), hence the value $\log 2$. However, should we employ a larger basis set, the value of $H_D$ would increase with each non-vanishing new natural orbital occupancy, and its values along the dissociation of $H_2$ would decrease as we reach the purely nondynamic correlation limit of two isolated one-electron fragments, which would give $H_D^\sigma=\log 2$ (hence, $H_D=H_D^{\alpha}+H_D^{\beta}=\log 4$ see Figure~\ref{fig:h2-dissociation-id-hd}).

It may be tempting to normalize the entropy by its formal maximum, \(\log(M)\), thereby defining 
\begin{equation}
\eta_{D}^{\sigma}=\frac{H_D^{\sigma}-\log(2)}{\log(M/2)}. 
\end{equation}
This quantity, conveniently bounded between 0 and 1 for non-idempotent matrices, measures the relative spread of $\{r_i^\sigma\}_{i=1\ldots M}$ distribution over the \(M\) available orbitals. The $M=2$ case would be singular for the reasons aforementioned. An analogous characterization is provided by \(\frac{N_{\mathrm{eff}}^{\sigma}-2}{M-2}\), which would approach a system-specific limiting value as \(M\) increases. Although conceptually appealing, neither \(\eta_D^{\sigma}\) nor \(\frac{N_{\mathrm{eff}}^{\sigma}-2}{M-2}\) is of much practical value: their convergence with \(M\) is slow, rendering them strongly basis-set dependent for the moderate basis sizes commonly employed in computational chemistry.\newline

The size dependence of the proposed indicator can
be established analytically for a collection of
noninteracting, identical fragments.
Consider $L$ identical fragments, each characterized
by the same set of orbital contributions and entropy
$H_{D,\mathrm{frag}}^\sigma$. Assuming that the natural
orbitals can be chosen as fragment-localized orbitals,
the normalized weights of the combined system are
$r_{i,\mathrm{frag}}^\sigma/L$ for each fragment.
The entropy of the complete system is therefore

\begin{align}
H_{D,L}^\sigma
&=-L\sum_i\frac{r_{i,\mathrm{frag}}^\sigma}{L}
\log\left(\frac{r_{i,\mathrm{frag}}^\sigma}{L}\right)\\
&=H_{D,\mathrm{frag}}^\sigma+\log L.
\end{align}

Thus the entropy is neither size-extensive nor
size-intensive. By contrast, its exponential satisfies
\begin{equation}
N_{\mathrm{eff},L}^\sigma
=L N_{\mathrm{eff},\mathrm{frag}}^\sigma.
\end{equation}
The effective number of contributing orbitals is therefore extensive and is consequently recommended as a more suitable measure of dynamic correlation.

\section{Computational Details}

All correlation indicators were computed from natural-orbital occupation numbers (NOONs) with an in-house Python~3 program (NumPy, SciPy and Pandas; Matplotlib for the figures). The NOONs were obtained by diagonalising the one-particle density matrix of each calculation. Following ref.~\citenum{xu:24jctc}, systems with $I_{ND}^{\max} > 0.030$ were classified as multireference (MR) and the rest as single-reference (SR). All calculations used closed-shell singlet references.

The following datasets were employed: 
\begin{itemize}

\item \textbf{GMTKN55 and AD3165.} 1327 closed-shell structures from the GMTKN55 database\cite{goerigk:17pccp} and the 3165 structures of the AD-3165 set,\cite{duan:20jctc} at the geometries of ref.~\citenum{xu:24jctc}. All 4492 structures were computed at the same level of theory, RHF followed by frozen-core MP2/cc-pVTZ, with PySCF.\cite{sun2018pyscf,sun2020pyscf} The cc-pVTZ-PP basis set and pseudopotential were used for elements beyond Kr, and def2-TZVP for K, for which no nonrelativistic cc-pVTZ basis set is available. The NOONs were obtained from the unrelaxed MP2 one-particle density matrix, which avoids the nonphysical occupations of the relaxed density, which is not necessarily $N$-representable. For AD-3165 this protocol reproduces the $I_{ND}^{\max}$ values of Ref.~\citenum{xu:24jctc}.

\item \textbf{Xtrem.} 31 small systems in three series covering extreme correlation regimes, computed at the full configuration interaction (FCI) level: the H$_2$ dissociation curve ($R = 0.5$--$4.0$~\AA{} and 10 and 12~\AA; FCI/cc-pVTZ with PySCF), and the helium (He to N$^{5+}$) and beryllium (Li$^-$ to O$^{4+}$) isoelectronic series. The isoelectronic series were computed at the all-electron FCI level with uncontracted even-tempered s, p, d basis sets (5 and 9 exponents, respectively), using Gaussian~03\cite{g03c02}, and the determinant-based FCI program of Knowles and Handy\cite{knowles:84cpl} and the DMn program\cite{dmn} to build the density matrix. 

\item \textbf{H$_2$ dissociation}. RHF and FCI calculations were performed with PySCF using the cc-pVDZ, cc-pVTZ, cc-pVQZ, cc-pV5Z and aug-cc-pVTZ basis sets, at 37 bond lengths between 0.5 and 4.0~\AA{} (steps of 0.1~\AA, plus $R_e = 0.735$~\AA).

\item \textbf{Basis set dependence} (Set A in Ref.~\citenum{xu:24jctc}) Seven small diatomic molecules at the equilibrium distance \(R_e\) and stretched geometry \(1.5 R_e\) studied by Handy et al.\cite{mok:96jpc}, computed with PySCF at the MP2 level of theory, employing various basis sets: 6-31G*, cc-pVDZ, cc-pVTZ, aug-cc-pVDZ, aug-cc-pVTZ. 

\item \textbf{H$_4$ D$_{2h}$/D$_{4h}$ pontential energy surface.} FCI calculations with the aug-cc-pVDZ basis set were performed with Gaussian~03~\cite{g03c02} and the FCI program by Knowles and Handy.~\cite{knowles:84cpl}

\end{itemize}

\section{Results}

\begin{figure}[H]
  \includegraphics[width=0.49\textwidth]{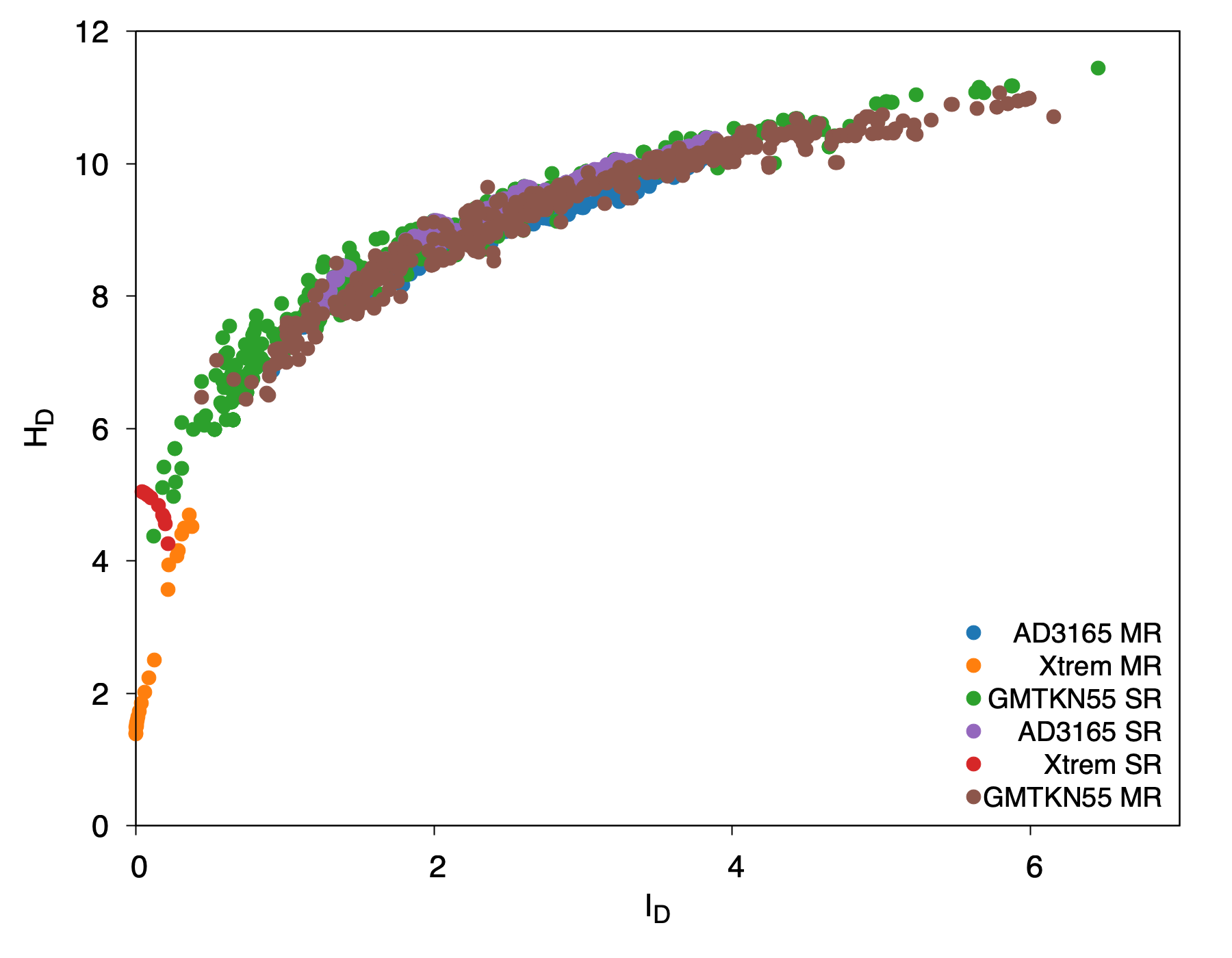}
  \hfill
  \includegraphics[width=0.49\textwidth]{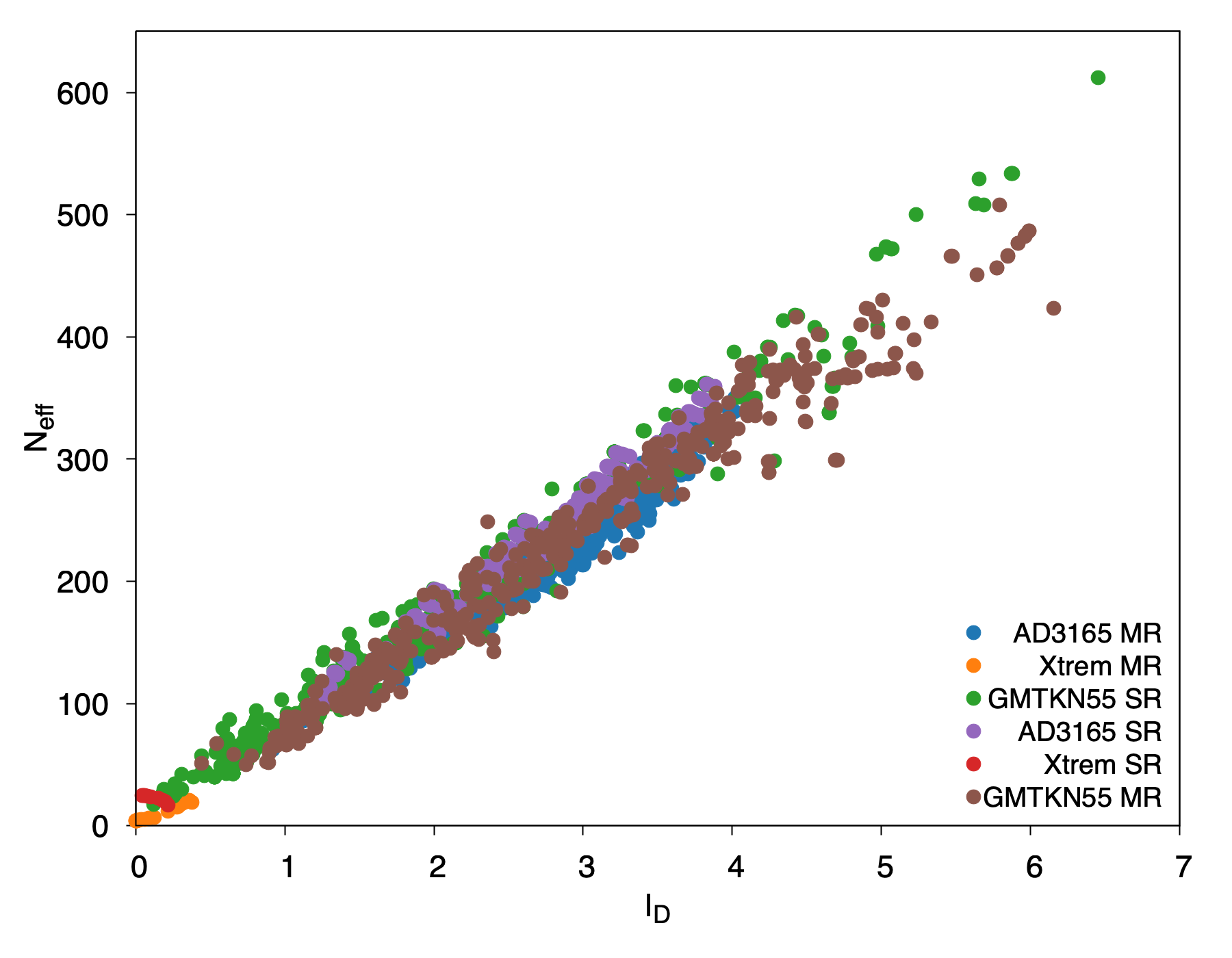}
  \caption{Relationship between \(H_D\) [left], its effective number of participating orbitals [right], \(N_{\mathrm{eff}}\), and the dynamic-correlation index \(I_D\).
  Colors distinguish the AD3165, Xtrem, and GMTKN55 sets, separated into single-reference (SR) and multireference (MR) subsets using $I_{\text{ND}}^{\text{max}}$ as a criterion.}
  \label{fig:hd-indicators}
\end{figure}

In Figure~\ref{fig:hd-indicators}, we have plotted $H_D$ and $N_{\mathrm{eff}}^\sigma$ vs. $I_D$ for the three datasets, divided into single-reference and multireference cases as classified by $I_{\text{ND}}^{\max}$.~\cite{xu:24jctc} The agreement between entropy-based indices and $I_D$ is excellent, confirming that both indices account for similar aspects of electron correlation. The almost-linear agreement between $N_{\mathrm{eff}}^\sigma$ and $I_D$, opens the room to connect $I_D$ to the effective number of correlated orbitals. It is particularly reassuring that the correlation stays among different datasets and among systems of different magnitude of nondynamic correlation, indicating that dynamic correlation is well-accounted for regardless the relevance of nondynamic correlation.\newline

\begin{figure}[H]
  \centering
  \includegraphics[width=\linewidth]{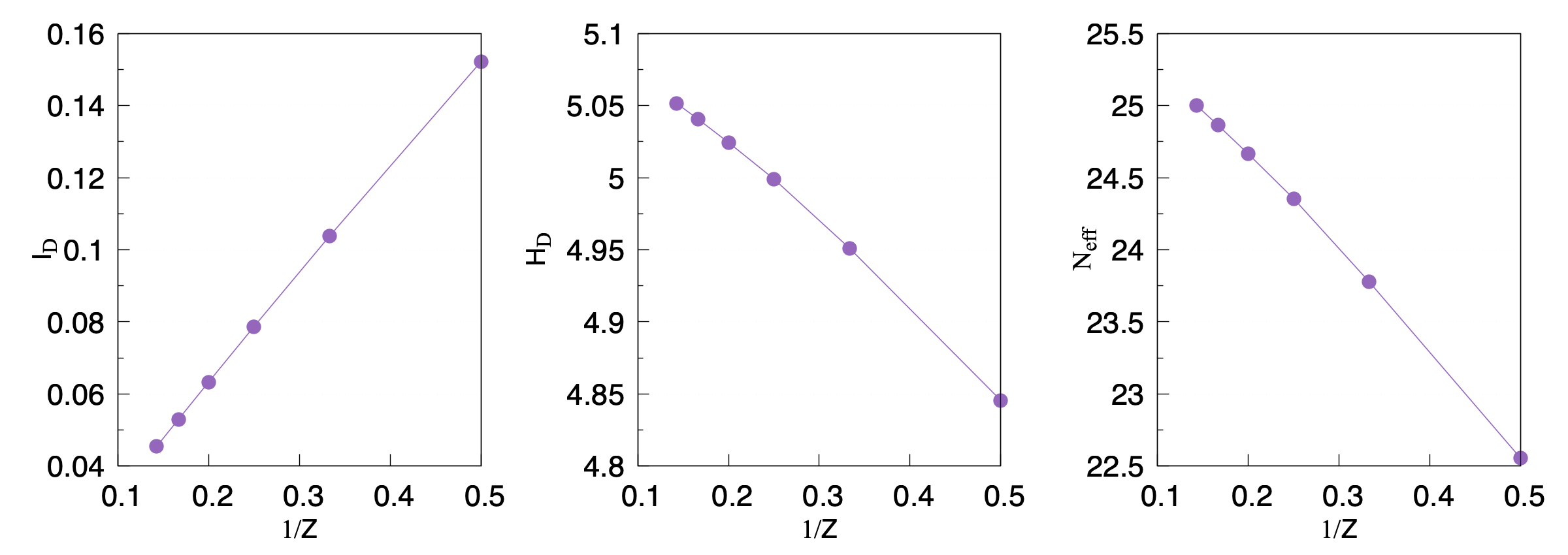}
  \caption{Indicators along the helium-like isoelectronic series as functions
  of $1/Z$: $I_D$ (left), $H_D$ (centre), and
  \(N_{\mathrm{eff}}\) (right).}
  \label{fig:he-isoelectronic-delta-hd}
\end{figure}

Despite their overall strong agreement, the two indices exhibit a few deviations for systems in the Xtrem SR set. Most notably, the He isoelectronic series displays opposite trends in the two descriptors, as shown in Figure~\ref{fig:he-isoelectronic-delta-hd}.

The He isoelectronic series is conventionally regarded as being dominated by dynamic correlation.\cite{ziesche:99jcp,katriel:06jcp,ramos-cordoba:16pccp} Its evolution along the series is, however, less straightforward to characterize. As \(Z\) increases, the electron--nucleus attraction increasingly dominates the electron--electron interaction, \(r_{12}^{-1}\). Accordingly, the leading spatial NOON approaches 2 and the total population of the \textit{correlation tail} ---the set of small occupations beyond the leading orbital--- decreases. This reduction in the magnitude of non-idempotency is captured by \(I_D\). At the same time, the normalized distribution of this tail becomes slightly more even among the remaining \(M-1\) orbitals. This broadening of the relative distribution explains the small increase of \(H_D\) with \(Z\).

Katriel and co-workers found that the densities and their scaled derivatives are remarkably similar throughout the He isoelectronic series.\cite{katriel:06jcp} Increasing \(Z\) therefore chiefly rescales and deepens an otherwise qualitatively unchanged Coulombic two-electron problem. Thus, it is questionable whether dynamic correlation changes significantly along the series.

The relatively large values of \(H_D\) and \(N_{\mathrm{eff}}^\sigma\) reflect the large number of \textit{effective} natural orbitals that carry the correlation tail. These quantities are not size-intensive: \(H_D\) has a logarithmic size dependence, whereas \(N_{\mathrm{eff}}^\sigma\) acts as an effective number of orbitals. After normalization by \(N\), \(N_{\mathrm{eff}}^\sigma/N\) places the entire He sequence among the systems with the largest effective number of correlated orbitals considered here. In contrast, \(\overline{I}_D=I_D/N\) identifies the low-\(Z\) members of the series as those with the largest amount of dynamic correlation (see Figure~S1).

The apparent disagreement is therefore not a contradiction, but a distinction between two aspects of correlation. If dynamic-correlation strength is identified with the depletion of the leading determinant and the total magnitude of the correlation tail, \(I_D\) is the more direct descriptor. If the focus is instead on the number and relative breadth of the \textit{effective} natural orbitals involved, \(H_D\) and \(N_{\mathrm{eff}}^\sigma\) provide complementary information. In most systems these two features increase together, explaining the strong correlations in Fig.~\ref{fig:hd-indicators}. The He isoelectronic series is thus a useful exception because radial scaling reduces the magnitude of the correlation tail while leaving its normalized distribution slightly more diffuse.\newline

\begin{figure}[H]
  \centering
  \includegraphics[width=1.0\linewidth]{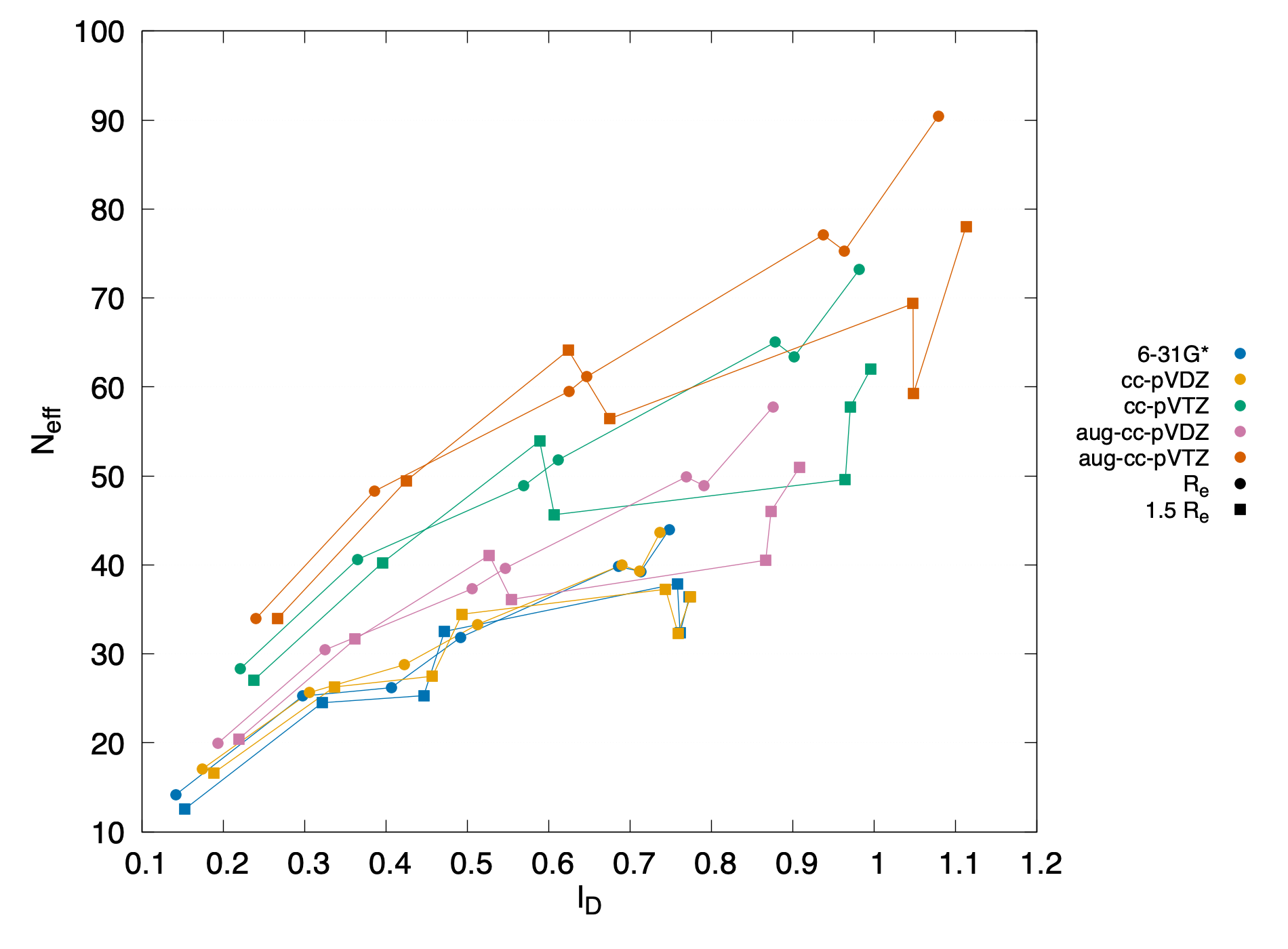}
  \caption{Basis-set dependence of \(N_{\mathrm{eff}}\) as a function of $I_D$. Circles and squares denote the
  equilibrium and stretched ($1.5R_\mathrm{e}$) geometries, respectively. Thin coloured lines serve only as guides to the basis-specific trends. The set consists of the diatomic molecules studied in Refs.~\citenum{mok:96jpc} and~\citenum{xu:24jctc}.}
  \label{fig:basis-dependence-delta-hd}
\end{figure}

As expected for descriptors of dynamic correlation, the indicators display an appreciable dependence on the one-particle basis, since the virtual orbital space determines how the correlated wavefunction can distribute its correlation tail (see Figure~\ref{fig:basis-dependence-delta-hd}). Nevertheless, the overall trends across the molecular set are reasonably well preserved: systems identified as more or less dynamically correlated retain broadly the same relative ordering across the basis sets considered.

Double-\(\zeta\) and, especially, 6-31G* bases provide an insufficient virtual space and therefore systematically underestimate the dynamic-correlation content accessible in larger bases. At least a triple-\(\zeta\)-quality basis set should consequently be used when the goal is to obtain a meaningful approximation to the dynamic-correlation signature of the wavefunction.\newline

\begin{figure}[H]
  \centering
  \includegraphics[width=0.90\linewidth]{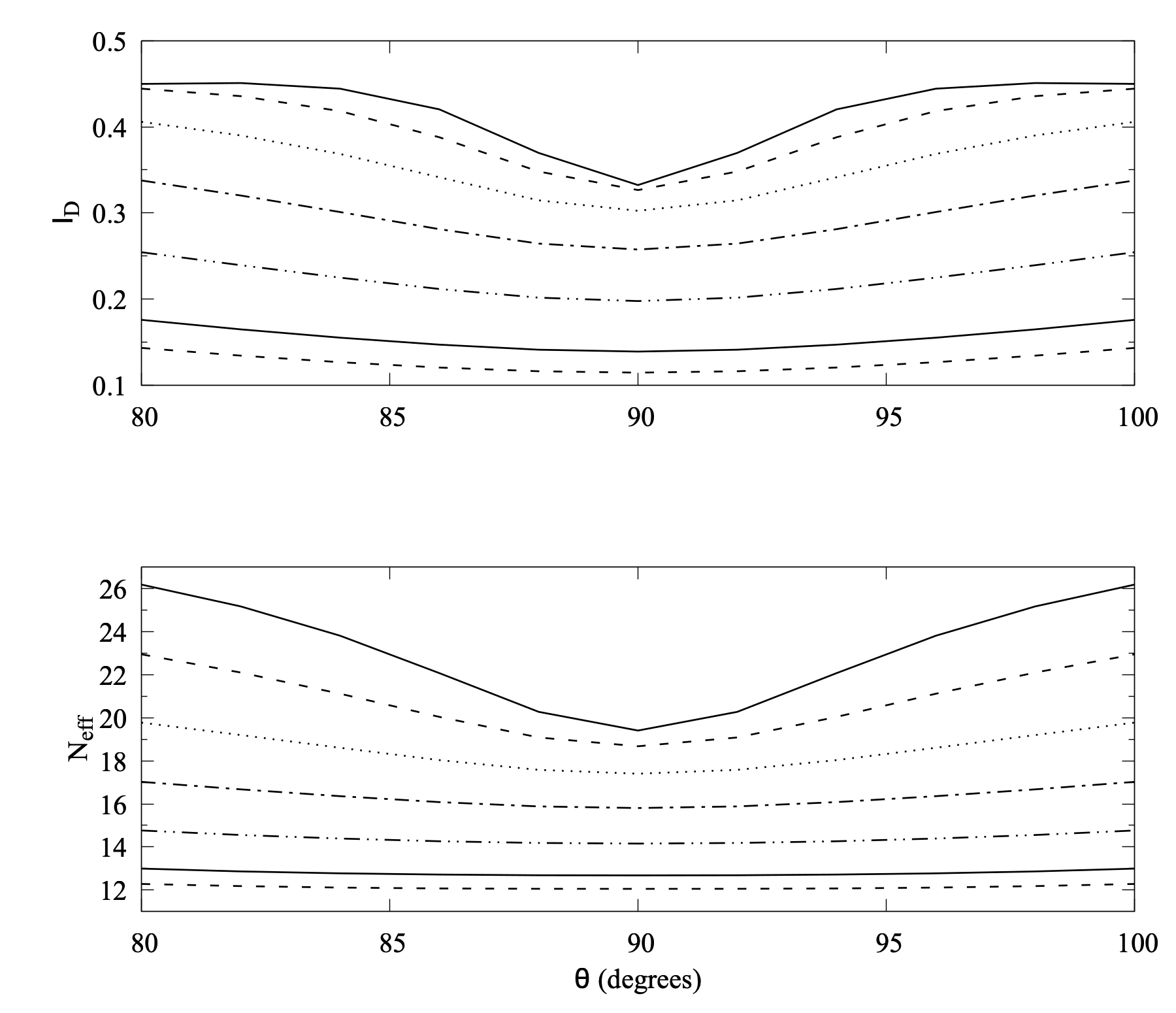}
  \caption{Dynamic-correlation regimes in the angular H$_4$ scan. Top:
  $I_D$; bottom: \(N_{\mathrm{eff}}\). Each curve corresponds to a fixed H--H distance
  $R$ between 0.8 and 2.0~\AA{}, with line styles progressing from the
  solid $R=0.8$~\AA{} curve to the most finely dashed $R=2.0$~\AA{} curve.}
  \label{fig:h4-delta-hd}
\end{figure}

Finally, we consider the singlet D$_{2h}$/D$_{4h}$ H$_4$ potential-energy surface~\cite{ramos-cordoba:15jcp} as an example in which geometrical changes shift the system from a dynamic-correlation-dominated regime to a nondynamic-correlation-dominated one.~\cite{ramos-cordoba:16pccp} To this end, we examine structures defined by different values of \(R\), the distance between each H atom and the center of mass, and \(\theta\), the angle formed by two adjacent H atoms and the center of mass. Increasing \(R\) stretches the H$_4$ molecule and enhances nondynamic correlation. In contrast, \(\theta\) controls the symmetry change from \(D_{4h}\) at \(\theta=90^\circ\) to \(D_{2h}\) upon distortion; thus, dynamic correlation increases as \(\theta\) departs from this value. Figure~\ref{fig:h4-delta-hd} confirms that both \(I_D\) and \(N_{\mathrm{eff}}\) follow the expected behavior.

\section{Conclusions}

We have introduced a natural-orbital occupation entropy, \(H_D^\sigma\), constructed from the normalized contributions of the natural orbitals to the total correlation measure \(I_{\mathrm{T}}^\sigma\). In contrast to indices that sum the magnitudes of the individual contributions, \(H_D^\sigma\) measures how these contributions are distributed among the orbitals. Its exponential, \(N_{\mathrm{eff}}^\sigma=\exp(H_D^\sigma)\), can consequently be interpreted as the effective number of natural orbitals participating in the correlation measure.

The proposed quantities display a strong overall correlation with \(I_{\mathrm{D}}\) for various molecular datasets covering both single-reference and multireference systems. This agreement supports the association between dynamic correlation and the participation of many weakly contributing orbitals. The He isoelectronic series illustrates that the entropy and \(I_{\mathrm{D}}\) nevertheless describe complementary features of the \textit{correlation tail}. Along this series, the magnitude of the tail decreases as \(Z\) increases, which is reflected by \(I_{\mathrm{D}}\), whereas its normalized distribution becomes slightly more diffuse, leading to a small increase in \(H_D\).

The \(\mathrm{H}_2\) dissociation and basis-set studies emphasize that \(H_D\) and \(N_{\mathrm{eff}}\) characterize the dynamic correlation represented by the wavefunction within the available orbital space. A sufficiently flexible basis set is therefore required to obtain a meaningful description of the correlation tail. In particular, double-\(\zeta\) and 6-31G* basis sets systematically underestimate the dynamic-correlation content accessible in larger bases, whereas triple-\(\zeta\)-quality basis sets preserve the qualitative trends more reliably and are therefore recommended as the minimum basis-set size for evaluating these descriptors.

\(H_D\) is neither size-intensive nor size-extensive. In contrast, \(N_{\mathrm{eff}}\) is extensive and is therefore recommended as the more useful quantity for assessing the importance of dynamic correlation. We anticipate that it may also serve as a target quantity for the development of more compact basis sets tailored to recover dynamic correlation and help identify local correlation domains for accelerated wavefunction methods based on orbital localization.

\begin{acknowledgement}

M.Y. acknowledges PIF2021 grant from Euskal Herriko Unibertsitatea. Grants PID2025-170025NB-C31 and PID2025-170025NB-C33 funded by MCIN/AEI/10.13039/501100011033 and ``FEDER Una manera de hacer Europa'', and the grants funded by the Gobierno Vasco (IT-2067-26) are acknowledged.
Calculations were performed on the computing facilities provided by the DIPC. Technical and human support provided by IZO-SGI, SGIker (UPV/EHU, MICINN, GV/EJ, ERDF, and ESF), and MCIA is gratefully acknowledged. E.M. acknowledges the Basque Government's BERC and the Severo Ochoa Centre of Excellence Programmes (CEX2024-001491-S, funded by MICIU/AEI/10.13039/501100011033).

\end{acknowledgement}

\section*{Data and Software Availability}

Calculated correlation indices, and data underlying all figures are available in machine-readable format in the Supporting Information. The scripts used to calculate the indices, process the data, and generate the figures are also provided in the Supporting Information.

\begin{suppinfo}

Additional computational details, supplementary figures, and data are available at \ldots

\end{suppinfo}

\bibliography{gen,selected_references}

\end{document}